\documentclass[aps,12pt,prd,superscriptaddress,preprintnumbers, 
	amssymb,
	amsmath,
	notitlepage,
	longbibliography,
	nofootinbib]{revtex4-1}
\usepackage[colorlinks=true, 
	linkcolor=blue, 
	citecolor=blue, 
	urlcolor=blue, 
	linktocpage=true]{hyperref}
\usepackage[utf8]{inputenc}
\usepackage[english]{babel}
\usepackage{lmodern}
\usepackage[protrusion=true, expansion=true]{microtype}
\usepackage[dvipsnames]{xcolor}
\usepackage{graphicx}
\usepackage{mathtools}

\usepackage[
]{todonotes}

\usepackage[framemethod=TikZ]{mdframed}
\usepackage{lipsum}

\mdfdefinestyle{MyFrame}{
linecolor=blue,
outerlinewidth=2pt,
roundcorner=20pt,
innertopmargin=\baselineskip,
innerbottommargin=\baselineskip,
innerrightmargin=20pt,
innerleftmargin=20pt,
backgroundcolor=gray!20!white}

\definecolor{remFcol}{RGB}{0,140,220}

\begin{document}

\title{Bubble nucleation at zero and nonzero temperatures}
\date{\today} 

\author{Mario Gutierrez Abed}
\email{M.L.Gutierrez-Abed2@newcastle.ac.uk}
\affiliation{School of Mathematics, Statistics and Physics, Newcastle University, 
Newcastle Upon Tyne, NE1 7RU, UK}
\author{Ian G. Moss}
\email{ian.moss@newcastle.ac.uk}
\affiliation{School of Mathematics, Statistics and Physics, Newcastle University, 
Newcastle Upon Tyne, NE1 7RU, UK}

\begin{abstract}
The theory of false vacuum decay in a thermal system may have a cross-over from
predominantly thermal transitions to quantum transitions as the temperature is
decreased. New numerical methods and results are presented here that can be used to
model thermal and vacuum bubble nucleation in this regime for cosmology and for 
laboratory analogues of early universe phase transitions.
\end{abstract}

\maketitle

\section{Introduction}

The early evolution of our universe is mostly a story of large scale homogeneity with small scale
perturbative fluctuations. Occasionally, though, non-perturbative effects may have played a role
during first order phase transitions. Characteristic features include the nucleation of bubbles,
possibly involving periods of extreme supercooling into a metastable, false vacuum state. Bubble
formation can be predominantly a quantum, or predominantly a thermal process. In this paper
we investigate the cross-over from thermal to vacuum nucleation in systems with
first order transitions.

Bubble nucleation in a thermal system can be described in terms of instantons,
solutions to an effective field theory with imaginary time coordinate
\cite{Coleman:1977py,Callan:1977pt,LINDE1983421}. The thermal aspect of
the decay is represented by imposing periodicity in the imaginary time coordinate,
with period $\beta=\hbar/(k_BT)$. At low temperatures, the size of the instanton
is small compared to $\beta$ and thermal effects appear mostly through the
form of the effective potential \cite{LINDE1983421}. At higher temperatures, 
provided the effective potential still has a potential barrier, the instanton solution becomes 
constant in the imaginary time direction. In between, there is a cross-over region were 
instanton solutions become distorted.

Interest in vacuum decay has been rekindled in the past few years by the possibility that
the process could be simulated in a laboratory Bose Einstein condensate 
\cite{FialkoFate2015,Braden:2017add,Braden:2018tky,Braden:2019vsw}. 
These systems will allow the first experimental tests of the theoretical
framework used to describe early universe phase transitions. 
It will be necessary to perform precise numerical modelling to compare theory with 
experimental results. Bubble nucleation
rates in cosmology are usually obtained using shooting methods 
(e.g. \cite{PhysRevD.45.2685,MEGEVAND201774}).
We will present a new numerical
method for calculating nucleation exponents for thermal vacuum decay 
applicable to the regime where both both thermal
and vacuum effects are important, and the shooting methods cannot be used.
This method can also be used when there is a background, or nucleation seed,
and it has already been used to obtain the results in Ref. \cite{Billam:2018pvp}, but the
method was not explained previously.

\section{The model}

We use a model based on the spinor BEC system of Fialko et al. \cite{FialkoFate2015},
where the relative phase between the wave-functions of two atomic states $\varphi$
is described by an action
\begin{equation}
S=\chi\int d^nxdt\left\{\frac12\dot\varphi^2-\frac12(\nabla\varphi)^2-V(\varphi)\right\}.
\label{action}
\end{equation}
Natural length and time units have been chosen based on the underlying physics
(explained later), and the parameter $\chi$ contains the remaining dependence on physical parameters.
The potential has been scaled to the form
\begin{equation}
V(\varphi)=-(1+\cos\varphi)+\frac12\lambda^2\sin^2\varphi.
\end{equation}
This potential has has two minima, a true vacuum at $\varphi=0$ and a false vacuum at $\varphi=\pi$,
separated by a potential barrier whose height depends on the parameter $\lambda$.
The number of spatial dimensions, $n$, depends on the details of the experiment, and
we will consider $n=1\dots 3$. The motivation for this potential is based on
a particular BEC system, but it also serves as a toy model for
early universe false vacuum decay, the essential features being that the 
system has a relativistic dispersion relation, and the potential
has the two minima separated by a potential barrier.

In a thermal system, the field responds to a modified potential that has 
$\lambda\equiv\lambda(T)$ \cite{Sher:1988mj}. In an early universe setting, this effect plays an
important role in placing the field in the false vacuum as the universe supercools.
In a laboratory setting, the phase is prepared in the false vacuum as part of the
experimental protocol. The potential barrier in the analogue model is still present at zero temperature,
and the temperature dependence of the potential plays far less of a role than it would in
some particle models. We will take $\lambda$ to be constant in the
modelling, and comment on temperature dependent parameters later.

The first-order false vacuum decay is a non-perturbative process,
in which quantum and thermal effects can contribute. In either case, the decay can 
be described by an instanton solution $\varphi_b$ to the field equations with imaginary time $\tau$. 
\begin{equation}
{\partial ^2\varphi\over \partial\tau^2}+\nabla^2\varphi-{\partial V\over\partial\varphi}=0
\end{equation}
In the vacuum case,
the field approaches the false vacuum value as $\tau\to\pm\infty$. In the
thermal case, an initial thermal ensemble is represented by solutions
that are periodic in $\tau$ with period $\beta=1/T$. We will refer to the special case
of an instanton solution which is independent of $\tau$ as a quasi-static instanton. 

The full expression for the nucleation rate of vacuum bubbles in a volume ${\cal V}$
depends on the  Euclidean action $S_E=iS$ of the instanton solution. According to
Coleman \cite{Coleman:1977py,Callan:1977pt},
\begin{equation}
\Gamma \approx {\cal V}\left|
{{\rm det}'\,S_E''[\varphi_b]\over {\rm det}\,S_E''[\varphi_{\rm fv}]}
\right|^{-1/2}\,\left({S_E[\varphi_b]\over 2\pi}\right)^{N/2}\,
e^{-S_E[\varphi_b]}.\label{gamma}
\end{equation}
where $S_E''$ denotes the second functional derivative of the Euclidean action, 
and det$'$ denotes omission of $N=n+1$ zero modes from the functional 
determinant of the operator in the vacuum case and $N=n$ zero modes
for the quasi-static instanton. The translational symmetry of the underlying
theory is broken by the instanton, and the zero modes are the modes representing
translations.

The action for a quasi-static instanton in one spatial dimension can be obtained analytically 
and provides a test for the numerical results we obtain later. In this case, 
the solution $\varphi\equiv\varphi(x)$ satisfies
\begin{equation}
{d^2\varphi\over dx^2}-{\partial V\over\partial\varphi}=0,
\end{equation}
with $\varphi\to\pi$ as $x\to-\infty$. This first integral of motion implies
$d\varphi/dx=(2V)^{1/2}$, and the solution bounces off the potential
at $\varphi_r=\arccos(1-2/\lambda^2)$. The action $S_E$ is
\begin{equation}
S_E=2\chi\beta\int_{\varphi_r}^\pi d\varphi\,(2V)^{1/2}.
\end{equation}
The integral can be obtained in closed form,
\begin{equation}
S_E=4\chi\beta\left\{(\lambda^2-1)^{1/2}-\lambda^{-1}\ln\left[(\lambda^2-1)^{1/2}+\lambda\right]\right\}
\label{Sexact}
\end{equation}
This exact solution is no longer valid in dimensions two and three, but it can be adapted,
for large $\lambda$, using the thin-wall approximation discussed below. 

The vacuum instanton in one spatial dimension has $O(2)$ symmetry, and
the solution is a function of $r=|{\bf x}|$. In the thin-wall approximation, the solution remains
close to the true vacuum value for small $r$, until a value $r\approx R$, when the solution
changes rapidly over a short distance (the `wall') with $d\varphi/dr\approx (2V)^{1/2}$. 
The Euclidean action in two dimensions can be approximated by splitting it up into
the interior and the wall,
\begin{equation}
S_E\approx -2\pi R^2\chi+4\pi \chi R\lambda
\end{equation}
There is an extremum at $R=\lambda$, where $S_E\approx 2\pi\chi\lambda^2$. For small
temperatures, this is lower than the quasi-static action form Eq. (\ref{Sexact}), 
$S_E\approx 4\chi\beta\lambda$.
In the thin wall approximation, vacuum tunnelling dominates at temperatures below 
$T\approx 2/(\lambda\pi)$, and thermal tunnelling dominates at higher temperatures.

The thin-wall approximation is only valid when the potential barrier is relatively large. 
Large barriers would be associated with bubble nucleation
rates too small to be relevant to cosmology or to be seen in the experiment. 
In the next section we look at new methods for evaluating the action that can go beyond 
the thin-wall approximation and give relevant nucleation rates.

The behaviour of the pre-factor in the nucleation rate (\ref{gamma}) can be analysed
by different numerical methods which we will not attempt to investigate here
\cite{Coleman:1985rnk,Garbrecht:2015yza,Ai_2019}. We note that,
since we are not in the thin-wall limit, there are no small parameters in the problem,
and so we expect the pre-factor to be of order one in the length and time units that
have been used in the action. Furthermore, the quasi-static and the vacuum
instantons approach one another at the cross-over from thermal to vacuum tunnelling, 
so the pre-factors will be the same at that point.

\section{Numerical method}

The instanton solution for false vacuum decay in $n$ spatial dimensions
has $O(n+1)$ symmetry, allowing the instanton equation to be reduced to an 
ordinary differential equation that is easily solved using shooting methods \cite{Coleman:1985rnk}. 
The reduced symmetry for the instantons in crossover regime of the thermal problem bars 
the use of this method. Although the instanton equations are a well-posed elliptic system, 
the negative and zero modes in $S_E''[\varphi_b]$ can be problematic for standard 
numerical techniques. We present a new relaxation method that overcomes these problems.

The basic relaxation method for solving a set of equations $S_E'[\varphi]=0$
introduces a field $\Phi$ that depends on ${\bf x}$, $\tau$ and a relaxation time $s$.
The field $\Phi$ solves
\begin{equation}
{d\Phi\over ds}=-{\cal O}S_E'[\Phi],\label{oldrelax}
\end{equation}
where the operator ${\cal O}$ is introduced to optimise convergence to the 
solution, $\Phi\to \varphi_b$ as $s\to\infty$. Close to the instanton solution,
the behaviour of $\Phi$ is governed by the second order operator $S_E''[\varphi_b]$.
If the solution to the relaxation equation is $\Phi=\varphi_b+\delta\varphi$, then the 
relaxation scheme for $\delta\varphi$ small reduces to
\begin{equation}
{d\delta\varphi\over ds}=-{\cal O}S_E''[\varphi_b]\delta\varphi.
\end{equation}
Choosing ${\cal O}$ so that ${\cal O}S_E''[\varphi_b]$ has a positive spectrum
leads to convergence in a neighbourhood of the solution. Since $S_E''[\varphi_b]$
has a negative eigenvalue, we cannot choose ${\cal O}$ to be a multiple of the identity.
The choice ${\cal O}=(S_E''[\Phi])^{-1}$ gives convergence, but it
requires a matrix inversion step that may itself be problematic due to small eigenvalues
of the operator.

A simple stability analysis by the von Neumann method shows that another obvious choice
${\cal O}=(S_E''[\varphi])^\dagger$ requires a very small 
numerical relaxation time step.  For a spatial step size $\Delta x$,
$\Delta s=O(\Delta x^4)$ for stability. 
However, this can be improved by taking a second order equation in the relaxation time,
\begin{equation}
{d^2\Phi\over ds^2}+2k{d\Phi\over ds}+(S_E''[\phi])^\dagger S_E'[\phi]=0,
\end{equation}
with a new parameter, the damping coefficient $k$. Using central differencing for the 
relaxation time derivatives, stability now requires $\Delta s=O(\Delta x^2)$.

The method works provided the initial guess for the bubble profile is sufficiently close 
to the final solution. In practice, a shape based on the thin-wall approximation serves
well. If the initial bubble radius is too small, then $\Phi$ relaxes to the false vacuum
state and a larger initial radius has to be selected.

The convergence of the method is related to the eigenvalue spectrum of $S_E''[\phi]$ .
If we consider a single mode with eigenvalue $\nu$, then the amplitude $\delta\varphi_\nu$
of the mode decays exponentially,
\begin{equation}
\delta\varphi_\nu\propto e^{-ks+(k^2-|\nu|^2)^{1/2}s}
\end{equation}
The zero modes are an exception, but the boundary conditions can be chosen to `pin' 
the centre of the instanton at the corner of the integration region to remove the 
(translational) zero modes. For large values of $|\nu|$, the convergence is determined by $k$, and for 
small $|\nu|$, by $|\nu|^2/(2k)$. The optimal value of $k$ would therefore be
$k\approx |\nu_{\rm min}|$, where $\nu_{\rm min}$ is the eigenvalue with smallest
modulus

\section{Results}

\begin{center}
\begin{figure}[htb]
\scalebox{0.2}{\includegraphics{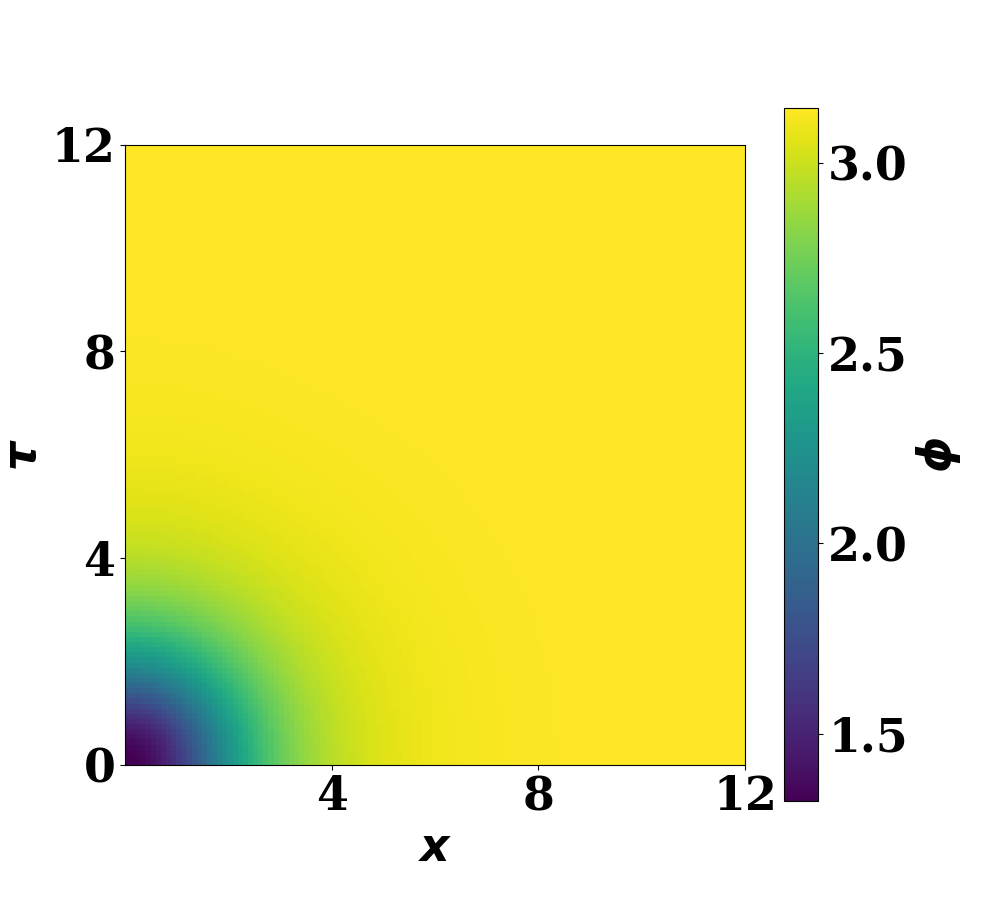}}
\scalebox{0.2}{\includegraphics{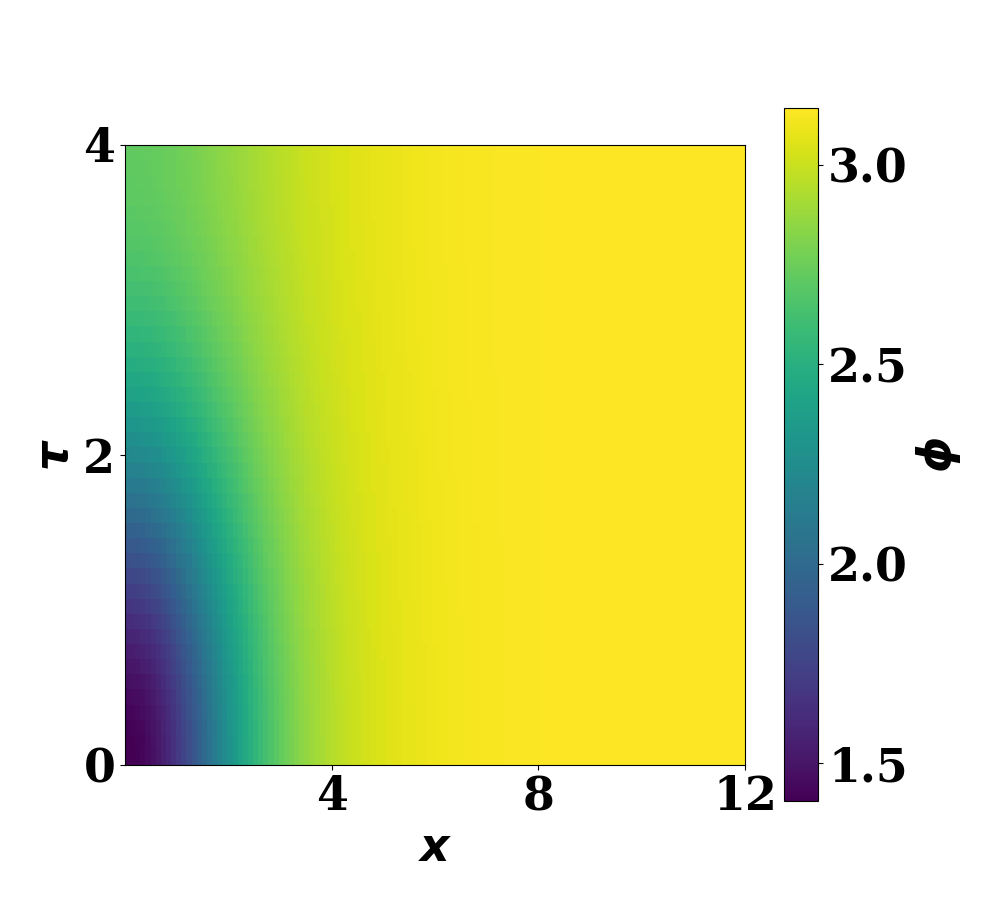}}
\scalebox{0.2}{\includegraphics{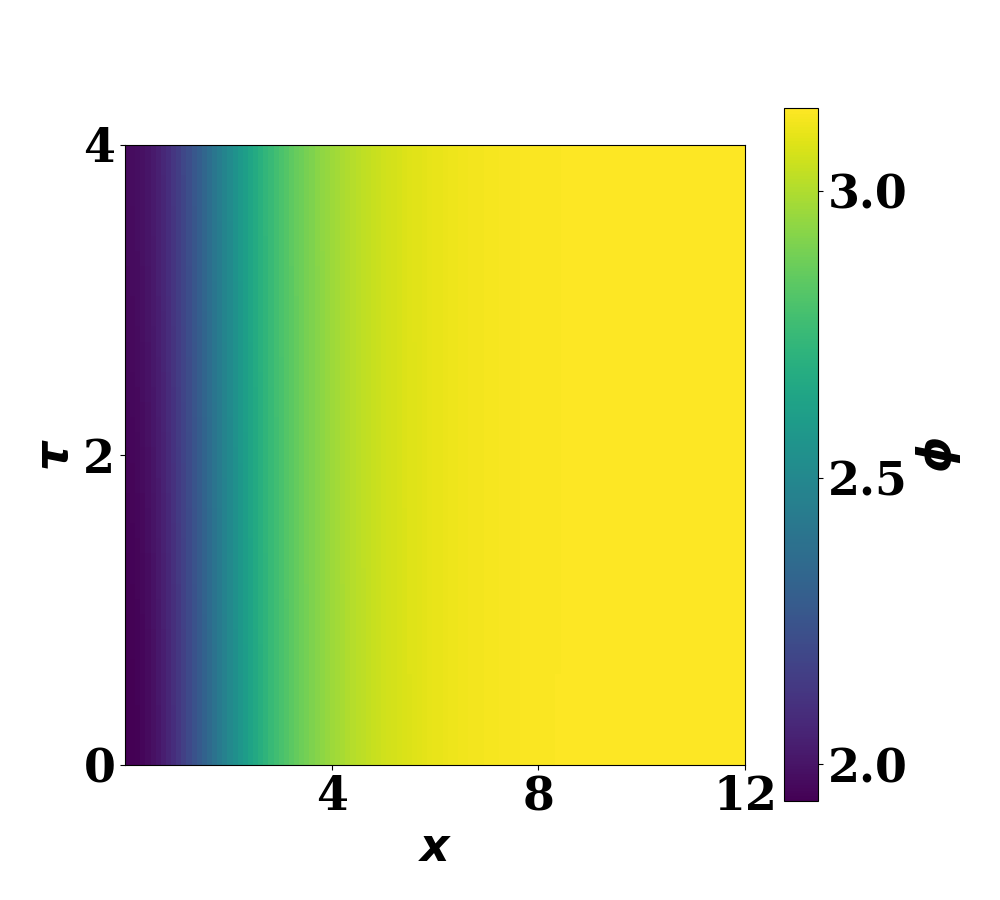}}
\caption{These plots show the value of $\varphi_b$ for shallow instantons in one dimension.
Left: The $O(2)$ symmetric vacuum instanton. Middle: The non-static instanton at $T=0.125$.
Right: The quasi-static instanton at $T=0.125$. In all cases $\lambda=1.2$.} \label{contour} 
\end{figure} 
\end{center}

Numerical results for the field of a non-static and quasi-static instanton solutions in
one dimension are  shown in figure \ref{contour}. At low temperatures, the non-static 
instanton approximates the $O(2)$ symmetric vacuum instanton. At higher temperatures,
in this case around $T=0.125$, the non-static instanton becomes distorted in the imaginary 
time direction. The quasi-static instanton solution is also shown. The radius
of the instantons in the spatial direction, defined as the distance to the average field value,
is between 2 and 3 length units.

Values of the Euclidean action at different temperatures are plotted in figure \ref{ST}. At low
temperatures, the non-static instanton has the lowest action. There is a crossover point where
the non-static solution merges into the quasi-static solution. We did not find any evidence for
a non-static solution with higher action than the quasi-static solution.

\begin{center}
\begin{figure}[htb]
\scalebox{0.5}{\includegraphics{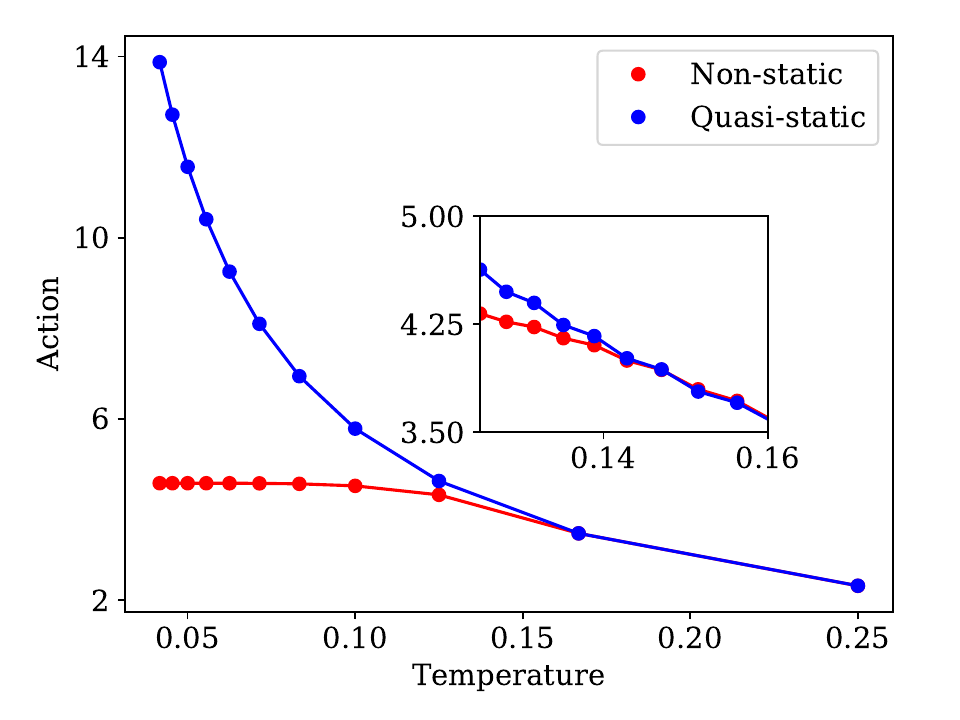}}
\scalebox{0.5}{\includegraphics{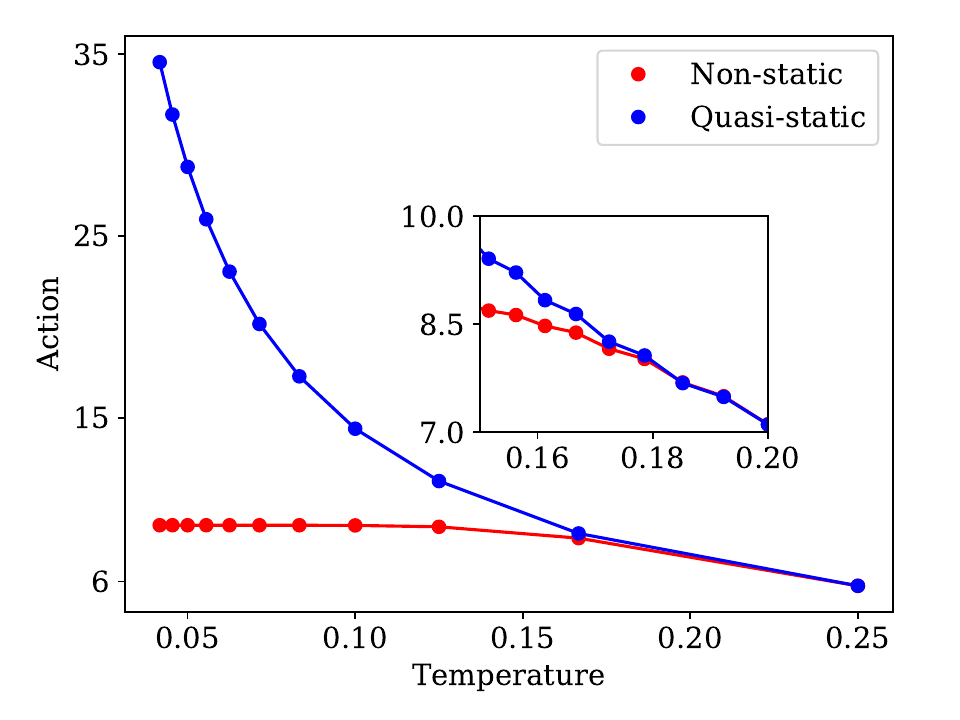}}
\caption{The plots show the dependence of the Euclidean action on temperature for non-static and
quasi-static instantons with different potential barrier heights. 
Left: $\lambda=1.2$; right: $\lambda=1.4$.} \label{ST} 
\end{figure} 
\end{center} 

\begin{center}
\begin{figure}[htb]
\scalebox{0.5}{\includegraphics{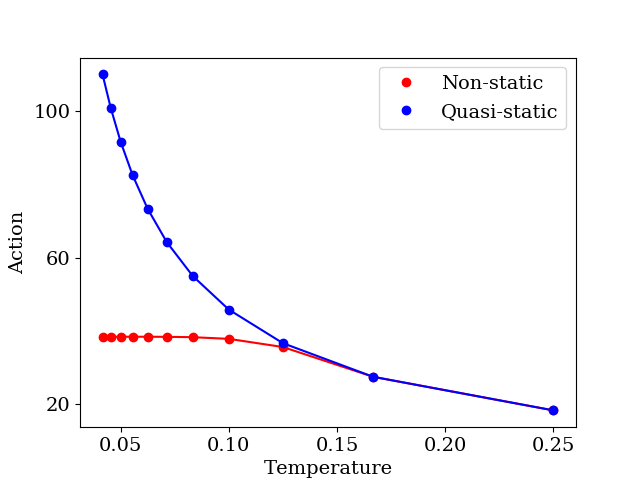}}
\scalebox{0.5}{\includegraphics{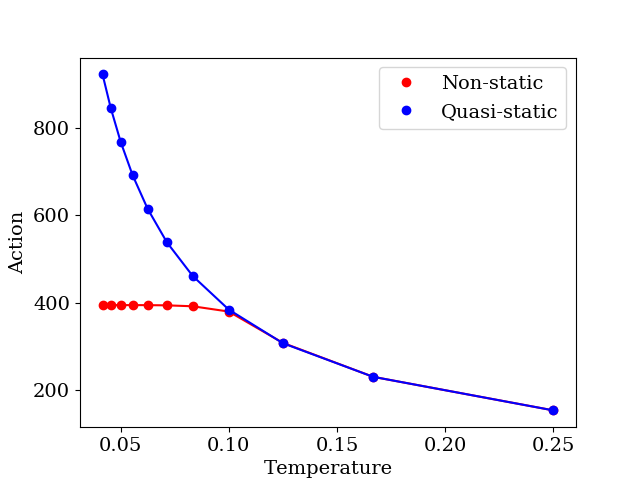}}
\caption{The plots show the dependence of the Euclidean action on temperature for non-static and
quasi-static instantons. Left:  two dimensions; right: three dimensions.
In both cases $\lambda=1.2$} \label{ST} 
\end{figure} 
\end{center} 

\begin{center}
\begin{figure}[htb]
\scalebox{0.45}{\includegraphics{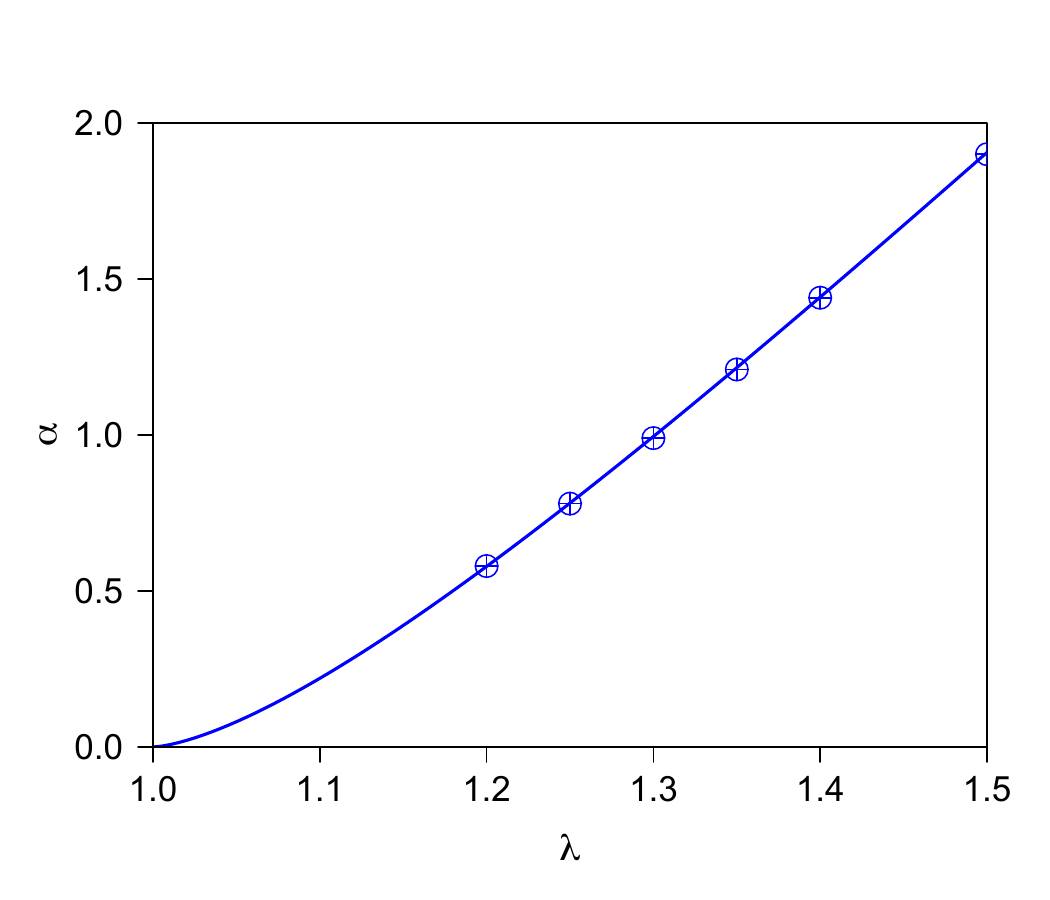}}
\scalebox{0.45}{\includegraphics{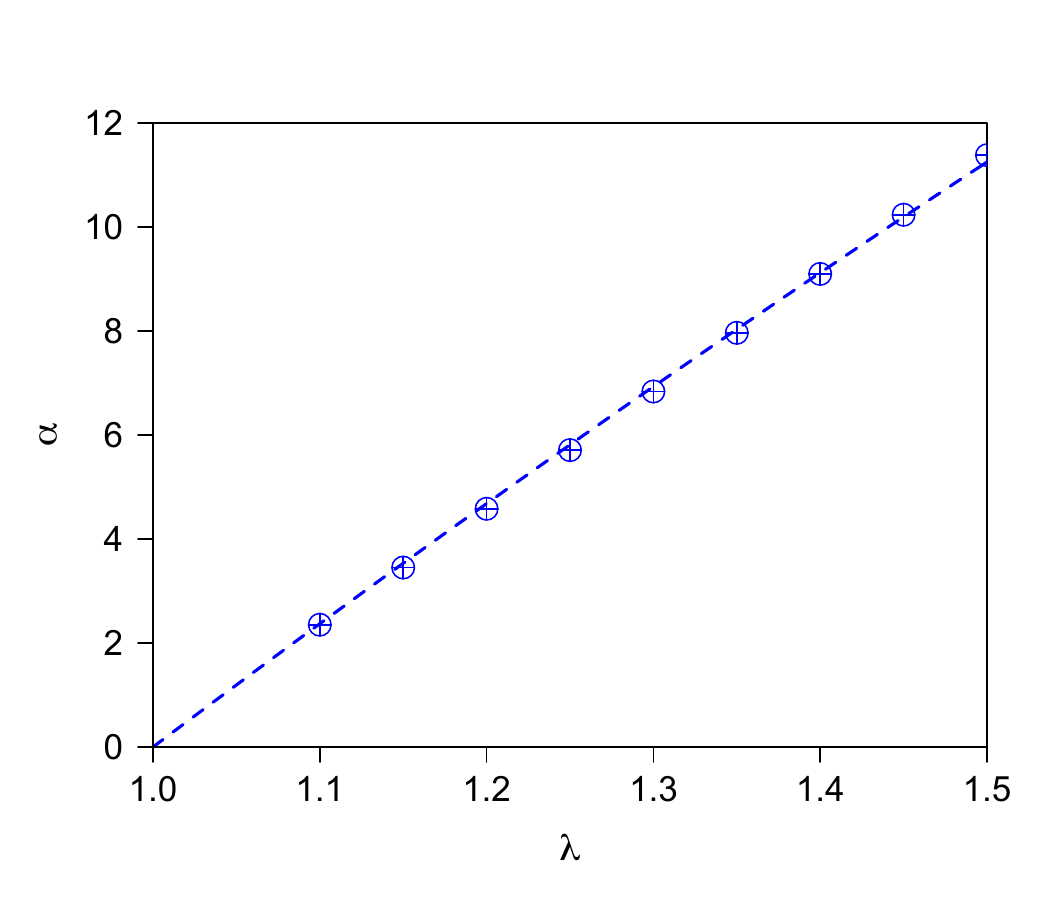}}\\
\scalebox{0.45}{\includegraphics{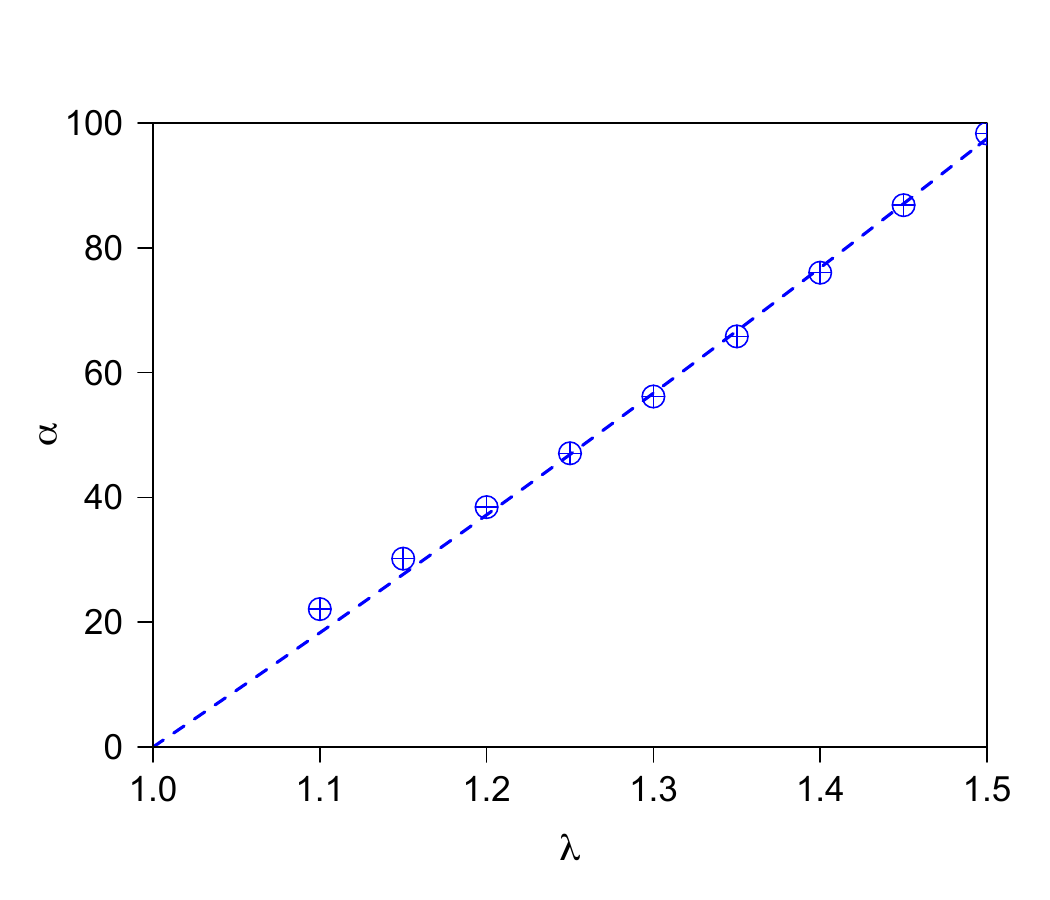}}
\caption{Values of $\alpha_n$ that determine the tunnelling exponent
for the quasi-static instanton in $S_E=\chi\alpha_n(\lambda)\beta$. The first plot is for
one dimension and also shows the exact analytic result using expression Eq. (\ref{Sexact}).
The second and third plots for two and three dimensions superimpose a numerical
fit $a_1(x-1)+a_2(x-1)^2$. In two dimensions $(a_1,a_2)=(24.0,-3.0)$ and in three dimensions
$(a_1,a_2)=(180,30.0)$}
\label{al}
\end{figure} 
\end{center}

In figure \ref{al}, we have taken the general form of the action for a quasi-static instanton
and parameterised this by
\begin{equation}
S_E=\chi\alpha_n(\lambda)\beta,\label{alphadef}
\end{equation}
in $n$ spatial dimensions. The vacuum instanton has action
\begin{equation}
S_E=\chi\alpha_{n+1}(\lambda),\label{alphadef}
\end{equation}
In one dimension, the agreement between the numerical 
values of $\alpha_n$ and the analytic expression Eq (\ref{Sexact}) is excellent. 
In two and three dimensions, the results differ substantially from the thin wall 
approximation. A numerical fit is shown instead.

\section{Conclusion}

We have investigated the cross-over regime of bubble nucleation
where the tunnelling instantons that dominate the nucleation rate
lose one degree of symmetry. The numerical results were obtained
using a new numerical method. We found that the distorted instantons 
merge smoothly into quasi-static instantons.

The results have been expressed in terms of a natural set of units
which are adaptable to the system under consideration.
In the case of the spinor gas, for example, the system has a characteristic
healing length $\xi$ and natural frequency $\omega_0$ \cite{FialkoFate2015}. 
The strength of the coupling between the spin states is tunable, and 
fixed by a small parameter $\epsilon$.
The units used for the numerical modelling are the length unit $\xi/(2\epsilon)$ 
and the temperature unit $2\hbar\omega_0\epsilon/k_B$.
The factor in front of the action (\ref{action}) in $n$ dimensions is
\begin{equation}
\chi=2^{-n}\epsilon^{1-n}\rho\,\xi^n,
\end{equation} 
where $\rho$ is the number density of atoms.
In the example from Ref. \cite{FialkoFate2015}, taking $5\times 10^5$ atoms of ${}^7{\rm Li}$ 
in a one dimensional atomic trap of length $120\mu {\rm m}$, the length unit would be 
$0.1\epsilon^{-1} \mu {\rm m}$ and the temperature unit $12\epsilon\,{\rm mK}$.

The analogue system has an asymmetric double well potential in the zero temperature limit.
There are models in particle physics with this behaviour, for example
the high-energy Higgs models used to discuss stability of the Higgs vacuum
\cite{Linde:1977mm,Sher:1988mj,Degrassi:2012ry,Branchina:2013jra,Branchina:2014rva}.
On the other hand, there are situations, such as variants of the standard model of particle
physics where the electroweak transition is first order, in which the potential barrier disappears
at zero temperature \cite{Kondo:1991jz,Kajantie:1996qd,ESPINOSA2012592}. 
Our numerical methods can only be extended to these situations
by taking into account the temperature dependence in the parameters
$\chi(T)$ and $\lambda(T)$. It would then be of possible to check, in each particular
model, whether the crossover between the different instanton types occurs
before the transition is completed.

\acknowledgments
IGM is supported by the Leverhulme Trust, grant RPG-2016-233, and
the Science and Facilities Council of the United Kingdom, grant number ST/P000371/1.
MGA is partially supported by a Newcastle Overseas Research Scholarship.

\bibliography{paper.bib}

\end{document}